\documentclass[slac_one]{revtex4}
\usepackage{graphicx}
\usepackage{fancyhdr}
\begin{document}

\title{Cosmological Constant in SUGRA Models Inspired by
Degenerate Vacua} 

%

\author{C. D. Froggatt, R. Nevzorov}
\affiliation{University of Glasgow, Glasgow, G12 8QQ, UK}
\author{H. B. Nielsen}
\affiliation{The Niels Bohr Institute, Copenhagen, DK 2100, Denmark}

\begin{abstract}
It is well known that global symmetries protect local
supersymmetry and a zero value for the cosmological constant in
no--scale supergravity. A particular breakdown of these symmetries, which
ensures the vanishing of the vacuum energy density, leads to the
natural realisation of the multiple point principle (MPP). In the
MPP inspired SUGRA models the cosmological constant is naturally
tiny.
\end{abstract}

\maketitle

\thispagestyle{fancy}


\section{INTRODUCTION}
Present astrophysical and cosmological observations indicate that
there is a tiny energy density (the cosmological constant)
$\Lambda \sim 10^{-123}M_{Pl}^4 \sim 10^{-55} M_Z^4$ spread all
over the Universe, which is responsible for its accelerated
expansion. This energy density is incredibly small as compared
with the contribution of QCD condensates to the total vacuum
energy density, which is expected to be of order
$\Lambda_{QCD}^4\simeq 10^{-74}M_{Pl}^4$. An even much larger
contribution $\sim v^4\simeq 10^{-62}M_{Pl}^4$ must come from the
electroweak symmetry breaking, while the contribution of
zero--modes should push the total vacuum energy density up to
$\sim M_{Pl}^4$. Because of the enormous cancellation between the
contributions of different condensates to $\Lambda$, the
smallness of the cosmological constant should be regarded as a
fine--tuning problem.

The smallness of the cosmological constant could be related to an
almost exact symmetry. An exact global supersymmetry (SUSY)
ensures zero value for the energy density at the minimum of the
potential of the scalar fields. However the breakdown of
supersymmetry induces a huge and positive contribution to the
total vacuum energy density of order $M_{S}^4$, where $M_{S}$ is
the SUSY breaking scale. The non--observation of superpartners of
quarks and leptons implies that $M_{S}\gg 100\,\mbox{GeV}$.

\section{NO-SCALE SUPERGRAVITY AND THE MULTIPLE POINT PRINCIPLE}
In $(N=1)$ supergravity (SUGRA) models the scalar potential is
specified in terms of the K$\ddot{a}$hler function
\begin{equation}
G(\phi_{M},\phi_{M}^{*})=K(\phi_{M},\phi_{M}^{*})+\ln|W(\phi_M)|^2\,,
\label{1}
\end{equation}
which is a combination of two functions: the K$\ddot{a}$hler potential
$K(\phi_{M},\phi_{M}^{*})$ and the superpotential $W(\phi_M)$. Here we
use standard supergravity mass units:
$\frac{M_{Pl}}{\sqrt{8\pi}}=1$. The SUGRA scalar potential is
given by
\begin{equation}
\begin{array}{c}
V(\phi_M,\phi^{*}_M)=\sum_{M,\,\bar{N}} e^{G}\left(G_{M}G^{M\bar{N}}
G_{\bar{N}}-3\right)+\frac{1}{2}\sum_{a}(D^{a})^2\,,\,\quad
D^{a}=g_{a}\sum_{i,\,j}\left(G_i T^a_{ij}\phi_j\right)\,,\\[2mm]
G_M \equiv \partial G/\partial \phi_M\,,\qquad
G_{\bar{M}}\equiv \partial G/ \partial \phi^{*}_M\,,\qquad 
G_{\bar{N}M}\equiv \frac{\partial^2 G}{\partial \phi^{*}_N \partial \phi_M}\,,\qquad
G^{M\bar{N}}=G_{\bar{N}M}^{-1}\,,
\end{array}
\label{2}
\end{equation}
where $g_a$ is the gauge coupling constant associated with the
generator $T^a$ of the gauge transformations. In order to break
supersymmetry in $(N=1)$ SUGRA models, a hidden sector is
introduced. It is assumed that the superfields of the hidden
sector $(z_i)$ interact with the observable ones only by means of
gravity. At the minimum of the scalar potential, hidden sector fields
acquire vacuum expectation values breaking local SUSY and generating
a non--zero gravitino mass $m_{3/2}=<e^{G/2}>$.

In general the vacuum energy density at the minimum of the SUGRA
scalar potential (\ref{2}) is huge and negative
$\rho_{\Lambda}\sim -m_{3/2}^2 M_{Pl}^2$, so that an enormous
fine--tuning is required to keep $\rho_{\Lambda}$ around the
observed value of the cosmological constant. The situation changes
dramatically if supergravity is supplemented by a global symmetry
that ensures a zero value for the cosmological constant. This is
precisely what happens in no-scale supergravity. The Lagrangian of
a no--scale SUGRA model is invariant under the imaginary
translations and dilatations which are subgroups of the $SU(1,1)$
group. The simplest no--scale SUGRA model involves only one hidden
sector superfield $T$ and a set of chiral supermultiplets
$\varphi_{\sigma}$ in the observable sector, which transform
differently under the imaginary translations ($T\to
T+i\beta,\,\varphi_{\sigma}\to \varphi_{\sigma}$) and dilatations
($T\to\alpha^2 T,\,\varphi_{\sigma}\to\alpha\,\varphi_{\sigma}$).
These symmetries constrain the K$\ddot{a}$hler function so that
\begin{equation}
K=-3\ln\biggl[T+\overline{T}-\sum_{\sigma}\zeta_{\sigma}|\varphi_{\sigma}|^2\biggr]\,,
\qquad W=\sum_{\sigma,\lambda,\gamma}\frac{1}{6}
Y_{\sigma\lambda\gamma}\,
\varphi_{\sigma}\varphi_{\lambda}\varphi_{\gamma}\,. \label{3}
\end{equation}
The structure of the K$\ddot{a}$hler function (\ref{3}) ensures
the vanishing of the vacuum energy density in the considered
model. Thus imaginary translations and dilatations protect a zero
value for the cosmological constant in supergravity. However these
symmetries also preserve supersymmetry in all vacua, which has to
be broken in any phenomenologically acceptable theory.

It was argued that the breakdown of dilatation invariance does not
necessarily result in a non--zero vacuum energy density
\cite{1}--\cite{Froggatt:2004gc}. Let us consider a SUGRA model with
two hidden sector fields $T$ and $z$. Suppose that $T$ and the
observable superfields $\varphi_{\sigma}$ transform under
imaginary translations and dilatations as before, whereas the
superfield $z$ transforms similarly to $\varphi_{\sigma}$.
Then the hidden sector superfield $z$ can appear in the full
superpotential of the model:
\begin{equation}
W(z,\,\varphi_{\alpha})=\displaystyle\kappa\biggl(z^3+ \mu_0
z^2+\sum_{n=4}^{\infty}c_n
z^n\biggr)+\sum_{\sigma,\beta,\gamma}\frac{1}{6}
Y_{\sigma\beta\gamma}\varphi_{\sigma}\varphi_{\beta}\varphi_{\gamma}\,.
\label{4}
\end{equation}
Here we include a bilinear mass term for the superfield $z$ and higher order
terms $c_n z^n$ that spoil dilatation invariance. At the same time we do not
allow the breakdown of dilatation invariance in the superpotential of the observable
sector, in order to avoid the appearance of potentially dangerous terms which lead,
for instance, to the so--called $\mu$--problem in the simplest SUSY models.

We also assume that the dilatation invariance is broken in the K$\ddot{a}$hler
potential of the observable sector, so that the full K$\ddot{a}$hler
potential takes the form:
\begin{equation}
K=-3\ln\biggl[T+\overline{T}-|z|^2-
\sum_{\sigma}\zeta_{\sigma}|\varphi_{\sigma}|^2\biggr]+\sum_{\sigma,
\lambda}\biggl(\frac{\eta_{\sigma\lambda}}{2}\,\varphi_{\sigma}\,
\varphi_{\lambda}+h.c.\biggr)+
\sum_{\sigma}\xi_{\sigma}|\varphi_{\sigma}|^2\,. \label{5}
\end{equation}
Here we restrict our consideration to the simplest set of terms
that break dilatation invariance in the K$\ddot{a}$hler potential.
Additional terms which are proportional to $|\varphi_{\alpha}|^2$
normally appear in minimal SUGRA models. The other terms
$\eta_{\alpha\beta}\varphi_{\alpha} \varphi_{\beta}$ give rise to
effective $\mu$ terms after the spontaneous breakdown of local
supersymmetry, solving the $\mu$ problem. We only allow the breakdown
of the dilatation invariance in the K$\ddot{a}$hler potential of
the observable sector, since any variations in the
K$\ddot{a}$hler potential of the hidden sector may spoil the
vanishing of the vacuum energy density in global minima.

In the considered SUGRA model the scalar potential of the hidden sector
is positive definite
\begin{equation}
V(T,\, z)=\frac{1}{3(T+\overline{T}-|z|^2)^2}
\biggl|\frac{\partial W(z)}{\partial z}\biggr|^2\,,
\label{6}
\end{equation}
so that the vacuum energy density vanishes near its global minima.
The minima of the scalar potential (\ref{6}) are attained at the
stationary points of the hidden sector superpotential. In the
simplest case when all the coefficients $c_n=0$, $W(z)$ has two extremum 
points at $z=0$ and $z=-\frac{2\mu_0}{3}$. In the first vacuum, where
$z=-\frac{2\mu_0}{3}$, local supersymmetry is broken so that
the gravitino becomes massive
\begin{equation}
m_{3/2}=\biggl<\frac{W(z)}{(T+\overline{T}-|z|^2)^{3/2}}\biggr>
=\frac{4\kappa\mu_0^3}{27\biggl<\biggl(T+\overline{T}
-\frac{4\mu_0^2}{9}\biggr)^{3/2}\biggr>}\,. \label{7}
\end{equation}
and all scalar particles get non--zero masses $m_{\sigma}\sim
\frac{m_{3/2} \xi_{\sigma} }{\zeta_{\sigma}}$. In the second
minimum, with $z=0$, the superpotential of the hidden sector vanishes
and local SUSY remains intact, so that the low--energy limit of
this theory is described by a pure SUSY model in flat Minkowski
space. If the high order terms $c_n z^n$ are present in
Eq.~(\ref{4}), the scalar potential of the hidden sector may have
many degenerate vacua with broken and unbroken supersymmetry in which
the vacuum energy density vanishes.

Thus the considered breakdown of dilatation invariance leads to
a natural realisation of the multiple point principle (MPP). The
MPP postulates the existence of many phases with the same energy
density which are allowed by a given theory \cite{Bennett:1993pj}.
When applied to $(N=1)$ supergravity, MPP implies the
existence of a vacuum in which the low--energy limit of the
considered theory is described by a pure supersymmetric model in
flat Minkowski space. According to the MPP this vacuum and the
physical one in which we live must be degenerate. Such a second
vacuum is only realised if the SUGRA scalar potential has a minimum
where the following conditions are satisfied
\begin{equation}
\biggl< W (z)\biggr>=\biggl<\frac{\partial W(z)}{\partial
z}\biggr>=0\,. \label{7a}
\end{equation}
This would normally require an extra fine-tuning. However, in the SUGRA models
considered above, the MPP conditions are fulfilled automatically
without any extra fine-tuning at the tree--level.

\section{COSMOLOGICAL CONSTANT IN THE MPP INSPIRED SUGRA MODELS}
Since the vacuum energy density of supersymmetric states in flat Minkowski space is
just zero and all vacua in the MPP inspired SUGRA models are degenerate, the
cosmological constant problem is thereby solved to first approximation by assumption.
However the value of the cosmological constant may differ from zero in the considered
models. This occurs if non--perturbative effects in the observable sector give rise
to the breakdown of supersymmetry in the second vacuum (phase). Our MPP philosophy
then requires that the physical phase in which local supersymmetry is broken in the hidden
sector has the same energy density as a phase where supersymmetry breakdown takes
place in the observable sector.

If supersymmetry breaking takes place in the second vacuum, it is caused by the strong
interactions. When the gauge couplings at high energies are identical in both vacua,
their running down to the scale $M_{S}\simeq m_{3/2}$ are also the same. Below the scale
$M_S$ all superparticles in the physical vacuum decouple and the corresponding beta
functions change. Using the value of $\alpha^{(1)}_3(M_Z)\approx 0.118\pm 0.003$
and the matching condition $\alpha^{(2)}_3(M_S)=\alpha^{(1)}_3(M_S)$, one finds 
the scale $\Lambda_{SQCD}$, where the supersymmetric QCD interactions become strong
in the second vacuum:
\begin{equation}
\Lambda_{SQCD}=M_{S}\exp\left[{\frac{2\pi}{b_3\alpha_3^{(2)}(M_{S})}}\right]\,,\qquad
\frac{1}{\alpha^{(2)}_3(M_S)}=\frac{1}{\alpha^{(1)}_3(M_Z)}-
\frac{\tilde{b}_3}{4\pi}\ln\frac{M^2_{S}}{M_Z^2}\,.
\label{8}
\end{equation}
In Eq.(\ref{8}) $\alpha^{(1)}_3$ and $\alpha^{(2)}_3$ are the values of the strong gauge
couplings in the physical and second minima of the SUGRA scalar potential, while
$\tilde{b}_3=-7$ and $b_3=-3$ are the one--loop beta functions of the SM and MSSM.
At the scale $\Lambda_{SQCD}$ the t--quark Yukawa coupling in the MSSM is of the same
order of magnitude as the strong gauge coupling. The large Yukawa coupling of the
top quark may result in the formation of a quark condensate that breaks supersymmetry
inducing a non--zero positive value for the cosmological constant
$\Lambda \simeq \Lambda_{SQCD}^4$.

\begin{figure*}[t]
\centering
~\hspace*{-7.5cm}{$\log[\Lambda_{SQCD}/M_{Pl}]$}\\
\includegraphics[width=100mm]{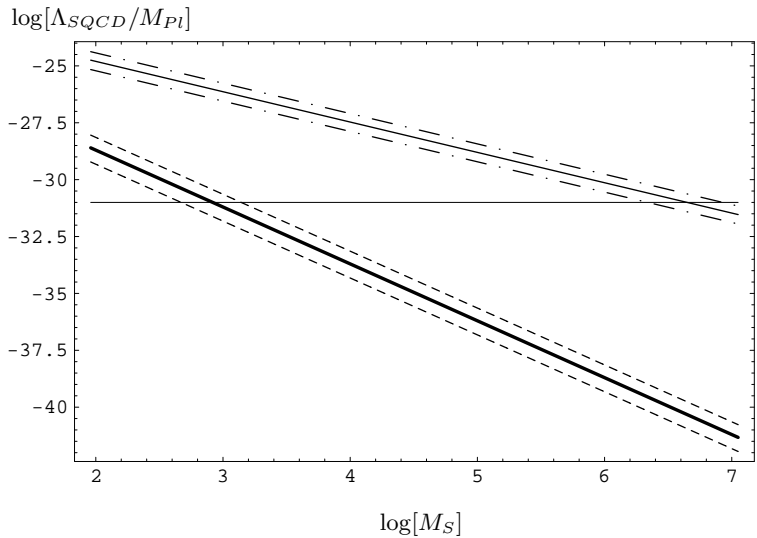}\\
~\hspace*{1cm}{$\log[M_S]$}\\
\caption{ The value of $\log\left[\Lambda_{SQCD}/M_{Pl}\right]$
versus $\log M_S$. The thin and thick solid lines correspond to
the pure MSSM and the MSSM with an extra pair of $5+\bar{5}$
multiplets. The dashed and dash--dotted lines represent the
uncertainty in $\alpha_3(M_Z)$, i.e. $\alpha_3(M_Z)=0.112-0.124$.
The horizontal line corresponds to the observed value of
$\Lambda^{1/4}$. The SUSY breaking scale $M_S$ is given in GeV. }
\label{lambda-qcd2}
\end{figure*}

In Fig.~1 the dependence of $\Lambda_{SQCD}$ on the SUSY breaking
scale $M_S$ is examined. Because $\tilde{b}_3 < b_3$ the QCD gauge
coupling below $M_S$ is larger in the physical minimum than in the
second one. Therefore the value of $\Lambda_{SQCD}$ is much lower
than the QCD scale in the Standard Model and diminishes with
increasing $M_S$. When the supersymmetry breaking scale in our
vacuum is of the order of 1 TeV, we obtain
$\Lambda_{SQCD}=10^{-26}M_{Pl} \simeq 100$ eV. This results in an
enormous suppression of the total vacuum energy density
($\Lambda\simeq 10^{-104} M_{Pl}^4$) compared to say an
electroweak scale contribution in our vacuum $v^4 \simeq 10^{-62}
M_{Pl}$.  From the rough estimate $\Lambda \simeq \Lambda_{SQCD}^4$
of the energy density, it can be
easily seen that the measured value of the cosmological constant
is reproduced when $\Lambda_{SQCD}=10^{-31}M_{Pl} \simeq 10^{-3}$
eV \cite{1}, \cite{Froggatt:2003jm}. The appropriate values of
$\Lambda_{SQCD}$ can therefore only be obtained for
$M_S=10^3-10^4\,\mbox{TeV}$. However the consequent large
splitting within SUSY multiplets would spoil gauge coupling
unification in the MSSM and reintroduce the hierarchy problem,
which would make the stabilisation of the electroweak scale rather
problematic.

A model consistent with electroweak symmetry breaking and
cosmological observations can be constructed, if the MSSM particle
content is supplemented by an additional pair of $5+\bar{5}$
multiplets. These new bosons and fermions would not affect gauge
coupling unification, because they form complete representations
of $SU(5)$. In the physical vacuum these extra particles would
gain masses around the supersymmetry breaking scale. The
corresponding mass terms in the superpotential are generated after
the spontaneous breaking of local supersymmetry, due to the
presence of the bilinear terms $\left[\eta (5\cdot
\overline{5})+h.c.\right]$ in the K$\ddot{a}$hler potential of the
observable sector. Near the second minimum of the SUGRA scalar
potential the new particles would be massless, since $m_{3/2}=0$.
Therefore they give a considerable contribution to the $\beta$
functions ($b_3=-2$), reducing $\Lambda_{SQCD}$ further. In this
case the observed value of the cosmological constant can be
reproduced even for $M_S\simeq 1\,\mbox{TeV}$ (see Fig.~1)
\cite{1}, \cite{Froggatt:2003jm}.

\begin{acknowledgments}
We would like to thank A. Faraggi, D.R.T. Jones, S.F. King, A.
Pilaftsis and J. Polchinski for fruitful discussions.

RN acknowledges support from the SHEFC grant HR03020 SUPA 36878.
\end{acknowledgments}

\end{document}